# ESG and the Cost of Capital:

# Insights from an AI-Assisted Systematic Literature Review[1]

Ebenezer Asem

Dhillon School of Business, University of Lethbridge, Canada

Email: ebenezer.asem@uleth.ca

Ruijie Fan

Dhillon School of Business, University of Lethbridge, Canada

Email: fanr@uleth.ca

Gloria Tian

Dhillon School of Business, University of Lethbridge, Canada

Email: gloria.tian@uleth.ca

---

[1] We thank comments and suggestions offered by Dr. Anastasia Stuart-Edwards, Dr. Piers Steel, Dr. Fan Yang, Dr. Lu Zhang and participants of the 2025 ASAC Conference. We also gratefully acknowledge financial support provided by a SSHRC Insight Grant. We used the AI-enabled platform HubMeta for data collection and quantitative analysis, ChatGPT-4o for editorial support, and Khairul K. Sumon also provided excellent research assistance. All remaining errors are our own.

**Abstract**

This paper explores how AI-powered tools could be leveraged to streamline the process of identifying, screening, and analyzing relevant literature in academic research. More specifically, we examine the documented relationship between environmental, social, and governance (ESG) factors and the cost of capital (CoC). By applying an AI-assisted workflow, we identified 36 published studies, synthesized their key findings, and highlighted relevant theories, moderators, and methodological challenges. Our analyses demonstrate the value of AI tools in enhancing business research processes and also contribute to the growing literature on the importance of ESG in the field of corporate finance.

# 1. Introduction

How does the practice of environmental, social, and governance principles (ESG) contribute to the sustainable development of the global economy? This question has gained increasing attention since 2004 when ESG principles were formally proposed and endorsed by the United Nations and heavy-weight players in the global financial sector (Li et al., 2021). With investors and policymakers placing greater emphasis on sustainability, understanding how ESG factors influence the costs of equity, debt, and overall capital is critical for companies aiming to balance financial performance with societal responsibilities. Furthermore, this relationship is particularly critical to understand during periods of economic uncertainty, such as the COVID-19 pandemic, which highlighted the importance of sustainable practices in mitigating risk and maintaining investor confidence.

Existing research has explored various dimensions of the ESG–cost of capital (CoC) nexus, including the role of various theories in explaining how ESG performance impacts financing costs. We are, however, unaware of systematic literature reviews that identify, evaluate, and integrate relevant studies on this important topic. This is understandable, given the methodological challenges such as synthesizing vast volumes of literature, addressing endogeneity concerns, and identifying key moderators across diverse contexts. These challenges are further exacerbated by the rapid growth of ESG research as well as the increasing complexity of financial markets. To address these challenges, we leverage artificial intelligence (AI) tools to enhance the efficiency and rigor of a systematic literature review on the association between ESG and CoC. These AI tools provide the ability to help create search terms, automate title screening, rank articles based on relevance, and extract and analyze data from large datasets, significantly reducing the time and effort required for manual reviews.

Our objectives are twofold: first, to synthesize the existing literature on ESG and CoC by identifying key theories, moderators, and methodological approaches; and second, to demonstrate how AI tools can support systematic literature reviews, offering a replicable framework for enhancing research efficiency and accuracy. We believe, by integrating AI into the research process, this study not only contributes to the substantive understanding of ESG-CoC relationship but also highlights the methodological advancement that AI offers to business research. The five specific research questions (RQs) we investigate are:

- RQ1: What statistical relationship between ESG and the CoC is documented in recent literature, especially when incorporating data from the COVID-19 pandemic period?
- RQ2: What are the key theories presented in recent studies on the ESG-CoC relationship?
- RQ3: Which moderating factors have been examined?
- RQ4: What instrumental variables (IVs) have been utilized to mitigate endogeneity?
- RQ5: What suggestions have been offered for future research in this area?

We are not the first to carry out a systematic literature review on ESG related themes. Prior systematic literature reviews have explored topics such as the motivations and consequences of ESG information, the characteristics of disclosure and its users that may influence outcomes (Tsang and Cao, 2023), the application of ESG scores (Clément and Trespeuch, 2023), and the theoretical basis of ESG research in broad scope (Li et al., 2021). Tsang and Cao (2023) concluded that investors and other stakeholders are shifting more attention to non-financial information (e.g. CSR or ESG) due to growing awareness of the importance of CSR/ESG practice in recent decades. When investors and other stakeholders assess the firms' ESG performance by comparing the ESG scores, ESG scores indicate the overall performance of environmental and social issues at a more macro level (Clément and Trespeuch, 2023). In addition, the previous literature review reveals the statistical information of keywords by frequency, which ranks the ESG-related keywords (Li et al., 2021). It shows that the most popular ESG topic is "corporate social responsibility" and it also manifests that stakeholder theory is one of the most important theoretical basis of ESG research (Li et al., 2021). The broad application of the stakeholder is also confirmed in section 3.2 of our systematic literature review paper. We are unaware of a systematic review of the relationship between ESG and the CoC, a relationship that has become an active topic of research in the face of COVID-19. We fill this gap in the literature and, equally importantly, we use this opportunity to document a replicable application of AI tools.

The remainder of the paper is organized as follows: Section 2 discusses in detail the AI-assisted methodology employed for our systematic literature review. Section 3 presents our findings on the ESG-CoC relationship, synthesizing insights from the 36 studies identified. Section 4 discusses the implications of AI-assisted research, offering some key takeaways and suggested directions for future research.

## 2. Methodology

Systematic literature reviews are important for synthesizing and analyzing existing research to support evidence-based decision-making. While these reviews have traditionally been conducted manually, the incorporation of AI tools is transforming the process by improving efficiency, scalability and accuracy (e.g. Alshami et al., 2023). This shift has been particularly noticeable in fields such as medicine and engineering. However, AI-assisted systematic literature reviews remain underutilized in business research.

The primary AI-enabled tool utilized in this paper is HubMeta (https://hubmeta.com/). On its homepage, Hubmeta is said to be "the first free web-based data entry system for correlational Meta-Analysis". It provides a user-friendly platform for uploading articles. Its filtering capability allows researchers to efficiently sort papers and automatically generate a preferred reporting items for systematic reviews and meta-analyses (PRISMA) diagram. It also includes the functionality to identify/control duplicates and record decisions researchers make during the review process. Hubmeta features a powerful machine-learning AI assistant/bot that, once enabled, learns from previous review decisions and assigns relevance scores to each article based on its fitness with the inclusion criteria. This AI bot can be applied to title and abstract screening as well as full-text screening processes. Regarding integrated analysis, Hubmeta platform offers a built-in meta-analysis tool that can generate complete correlation matrices, streamlining the synthesis process.

**2.1 Collecting Potentially Relevant Articles**

In an AI-assisted systematic review, selection of search terms to extract relevant research articles is an important first step, and there is a fine balance between keeping the search terms broad and yet, ensuring that they are specific and focused. We, the research team, started by putting together our own lists of potential keywords, relating to ESG and Cost of Capital (CoC) respectively. We then used ChatGPT-4o to generate mirroring lists of keywords and merged the two sets of keyword lists. Lastly, ChatGPT-4o was used to ensure that our final list of search terms conform to the search code style requirement of our chosen academic search engine, Web of Science. This was a valuable starting point, as Web of Science offers extensive and well-curated data for journal articles and conference proceedings (Steel et al., 2023). We extracted all articles from Web of Science that contained the chosen search terms in the article title or abstract. We also restricted publications from 2020 to 2023, given that we are interested in examining the ESG-CoC relationship in the presence of the recent pandemic[2]. Our initial search step yielded 876 articles. The PRISMA flowchart illustrating this step, along with subsequent article collection steps, is provided in Figure 1. Our final search codes with keywords are presented in Figure 2.

The titles and abstracts of these 876 articles, compiled into one .RIS file, were then uploaded to the Hubmeta platform. This platform developed an algorithm that can automatically compare the information about article title, authorship, year, publisher and abstract among the papers uploaded and to report article duplications. Implementing such algorithm helped us reduce the total number of articles from 876 to 869.

## 2.2 Title and Abstract Screening

The second step is to screen article titles and abstracts. Although this process has traditionally been labour-intensive, the use of HubMeta made it more manageable. To minimize selection bias, we started the screening process using two research assistants. One assistant is a junior researcher from the research team, and the other assistant is a well-trained finance master's student who is unaffiliated with the team. The two research assistants independently evaluated 50% of the collected articles (i.e. 435 out of 869) based on pre-specified inclusion criteria. Any discrepancies between their decisions were then referred to the project managers, the senior researchers of the research team, for further discussion and resolution. Articles were included if they met all of the following criteria:

- The study is relevant to COVID-19, and its data encompasses at least a period between 2020 and 2022 (i.e. "the COVID years").

---

[2] *This first step of collecting research articles was completed on August 25th, 2023 and, as such, papers published after this date are not included in our systematic literature review.*

- The response variable relates to at least one of cost of capital/cost of equity/cost of debt.
- The primary explanatory variable(s) is/are related to the Refinitiv ESG or MSCI ESG definitions. We present the relevant ESG frameworks in Figure 3 and Figure 4.

Decisions and comments made by the two research assistants were recorded in HubMeta. The platform's AI bot learned from these decisions and then ranked the articles based on their relevance to the inclusion/exclusion criteria. Specifically, the HubMeta AI bot provided scores on the articles, driven by its analysis of all previous decisions. The scores range from 0 to 1, with increments of 0.01, representing probabilities calculated using a similarity assessment algorithm employed by the AI tool. For example, a score of 0.88 suggested an 88% probability that the article would meet our inclusion criteria. The AI bot sorted all articles in descending order of their scores, with higher-scored articles appearing first. As more articled were being processed and ranked, the scores of the remaining articles tended to decrease over time. This helped streamline the screening process, enabling us to complete this step more efficiently than using a manual approach. To train the AI bot, we used "tfidf" as our Feature extraction model and used "logistic" as our classifier model. Saeidmehr et al., (2024) highlighted that the spiral processing approach with logistic regression, TF-IDF for vectorization, and maximum probability for prioritization demonstrated up to a 90% improvement over traditional machine learning methodologies.

A reliable machine learning stopping rule is to identify a specific cut-off score below which one would observe consistent manual rejection of articles. The AI bot sorts the articles in descending order based on their scores. The researcher monitors how many articles are consecutively rejected, which helps determine at what time the researchers should stop reviewing further articles. For our study, we noticed that articles with scores below 0.53 were consistently rejected, indicating that the remaining articles in our database, evaluated by the AI bot to have scores below 0.53 (i.e. our stopping rule), would unlikely be relevant. Noroozi et al. (2023), in their AI-assisted systematic review on the impact of vehicle automation, highlighted the importance to control the errors of validating the classification model. To test our type II error, we investigated how many papers were assigned a relevance score lower than 0.53 but were accepted manually. We verified the Type II error rate in our database, which was less than 5 percent, indicating that the HubMeta AI bot had a very low likelihood of excluding a relevant article from our project dataset. Working as a qualified and experienced researcher with a low type II error, the AI assistant help save 50% of time in the title and abstract screening stage. Our AI-assisted title and abstract screening process excluded 589 papers, retaining 280 research articles for full-text screening.

**2.3 Full-text Screening**

The final step of article collection is to conduct full-text screening of the articles selected after title and abstract screening. As the number of articles requiring full-text screening was manageable, the two research assistants performed this step entirely manually to ensure accuracy. This process yielded 36 research articles, all peer-reviewed, for our subsequent systematic literature review.

**3. Main Findings**

Table 2 summarizes the 36 research articles, and they are categorized by market regions and across the three cost of capital themes – cost of equity, cost of debt, and the weighted average cost of capital. Themes wise, we observe 30.56% (11 of 36) of these studies analyzed the cost of equity. The Asian markets emerged as a primary focus, with 5 studies exploring the theme of cost of equity. Two studies investigating European markets are included in the systematic literature review paper.

African and Middle Eastern markets had minimal representation, with 1 paper each within the cost of equity theme. The "multiple markets" (i.e. international) category featured 3 studies, indicating cross-regional evidence to support the benefit of sustainable development. The cost of debt was the most extensively studied theme, with 24 research articles in total. The Asian market led again with 11 studies, manifesting the research interest in debt-related business sustainability. The European market followed with 5 studies, while the American and "multiple markets" category contributed 2 and 5 studies respectively. The Middle Eastern region had the least representation, with only 1 study included in this systematic literature review. The Weighted Average Cost of Capital (WACC) received minimal research coverage, with just 1 paper addressing the relationship between sustainability and the WACC. This sole study focused on the Asian market, leaving a research gap for exploration for other regions.

The study includes the following as its first inclusion criterion: The study is relevant to COVID-19, and its data encompasses at least a period between 2020 and 2022 (i.e., the COVID-19 years). As COVID was not included as a theme (the themes focus on the cost of equity, cost of debt and the weighted average cost of capital), we show examples of how the selected papers are related to the pandemic. As shown in Table 1, the 36 papers are divided into two groups: the high-relevance group (9 papers, 25%) and the medium-relevance group (27 papers, 75%). The papers in the high-relevance group have a stronger tie with COVID, including 3 papers on the cost of equity and 6 papers on the cost of debt. These papers in high-relevance groups discuss the side effects of COVID-19 on the cost of capital and the insurance-like mechanism of superior ESG performance. Srivastava (2022) investigates whether ESG is the key to unlocking debt financing during the COVID pandemic, with data from January 2016 to December 2020. A COVID dummy is constructed to capture the impact of the pandemic. The dummy variable interacts with the ESG variable to capture the ESG premium in the regression. The results show that firms with greater stakeholder engagement obtain higher debt financing during the COVID-19 pandemic, especially for firms with higher risk levels and lower asset-intensive levels (Srivastava et al., 2022). Creating the dummy and interacting with ESG-related variables is also employed in other studies (Agnese and Giacomini, 2023; Lian et al., 2023). Alternatively, other studies estimate the regression model before and after COVID-19 (Chouaibi et al., 2022), or estimate within the pandemic period (Ferriani, 2023), or estimate by re-running the regression with pandemic year added (Ali et al., 2022). In table 1, the papers in the medium-relevance group have a tie with COVID but do not involve a detailed discussion regarding the impact of COVID or the impact of ESG benefits during the pandemic. Their study's data encompasses a period between 2020 and 2022 (i.e., the COVID years). Even though they do not explicitly focus on COVID-19 or ESG benefits during the pandemic, these studies may still reflect indirect impacts, such as whether the relationship between sustainability practices and cost of capital holds with the incorporation of the COVID periods.

## 3.1 Relationship between ESG and CoC

We conducted a meta-analysis to explore the relationship between ESG and the cost of capital. Figure 5 shows how ESG and cost of capital variables were constructed. Using HubMeta's image processing capabilities, we extracted data from correlation tables, ensuring accuracy through thorough manual reviews to resolve any inconsistencies. The Hedge's Q test of homogeneity is 0.1344 (greater than 0.05), indicating that all studies share a common effect size is not rejected. This highlights that results from different studies are consistent enough to be pooled together. Applying the composite scores technique (Hunter, 2004), we found a negative correlation of -0.1197 (excluding the regression coefficient), statistically significant at the 5% significance level. This result indicates that, on average, companies with stronger ESG performance benefit from lower costs of capital, likely because investors and lenders perceive such companies as lower-risk or more sustainable, thereby reducing their financing costs. This finding directly addresses our first research question (RQ1).

## 3.2 Summary of applicable theories

Literature on the relationship between ESG and the CoC spans across several popular theories, including the risk-reward theory, agency theory, stakeholder theory, signalling theory, trade-off theory, decision-usefulness theory, legitimacy theory, volunteer disclosure theory, resource dependent theory, and peer effect theory. Among the 36 articles reviewed, many theories were applied in both cost of equity papers and cost of debt studies and the underlying logic of an applied theory is largely consistent across these two themes. Among these, risk-reward theory, agency theory, and stakeholder theory are the most frequently applied, appearing in the majority of the 36 articles. Therefore, this section focuses on discussing these three predominant theories. It addresses our second research question (RQ2), with a more comprehensive summary of all theoretical frameworks presented in Table 3.

**Stakeholder Theory**

Stakeholder theory posits that enterprises should look beyond the maximization of short-term shareholder profits by considering the impact of their operations on the interests of all stakeholders, including but not limited to the community, society and the environment (Freeman, 1984). Chouaibi (2021) proposed that corporate social responsibility should have a negative effect on the cost of equity. Tang (2022) also claimed that ESG performance to be negatively related to an enterprise's cost of equity. Stakeholder theory can rationalize the ESG and firm value as ESG increases shareholder wealth, because it motivates other stakeholders to contribute to the success of the firm with the resources they avail (Freeman, 1984; Freeman, 2010). Supported by stakeholder theory, firms with higher ESG are shown to be less likely to face lawsuit risk due to the reduction of environmental pollution (Sharfman and Fernando, 2008). Lynch and O'Hagan-Luff (2023) proposed that the relationship between corporate social performance (CSP) and the cost of equity is stronger in the presence of stakeholder-supporting institutions. In aggregation, investors take an instrumental view of CSP, pricing it based on its implications for firm risk and performance (Garriga & Melé, 2004), and that institutional structures alter its associated costs and benefits by altering stakeholder salience (Mitchell et al., 1997). Higher stakeholder salience, driven by stronger stakeholder-supporting institutions, enhances the perceived value of CSP, as failing to address the concerns of salient stakeholders increases the risk of suboptimal financial outcomes for the firm.

**Agency Theory/Information Asymmetry**

Corporate governance has a significant positive impact on both the firms' cost of equity capital and cost of debt capital. Conflicts of interest between principals and agents may incur cost for business. Monitoring cost is one of the agency costs that most companies undertake. Corporate governance is believed to reduce agency costs in the framework of agency theory/information asymmetry and in COEs by reducing the risks observed by investors (Lambert et al., 2007). According to Zandi et al. (2020), an enhancement in cash would increase the corporate agency cost so directors could apply cash flows to their interests than corporate shareholders.

**Risk-Reward Theory**

Risk-reward theory explains that firms are motivated to pursue sustainable performance because potential long-term rewards of sustainable practices could outweigh the associated risks. It explains the benefit of ESG practice from the perspective of risk reduction. Integrating non-financial factors in the investment process could lead to both a higher profitability and to better risk management (e.g. Galema et al., 2008; Lins et al., 2017; Agnese & Giacomini, 2022). The reduction of risk is considered a core driver for sustainability, which is related to contingencies, potential and actual costs, that can influence future economic and business performance (Schaltegger and Burritt, 2018). Godfrey et al. (2009) provide evidence that managers of firms who engage in sustainability can create value at times for their shareholders by creating insurance-like protection. By adopting sustainable practices, firms reduce their exposure to risks such as regulatory penalties, resource shortages, and reputational damage. Risk-reward theory highlights that the costs of inaction are high, motivating firms to invest in sustainability to avoid these potential losses. For example, effective management of ESG issues facilitates better risk management practices, which may impact both the probability of risk ex-ante and the severity of losses ex-post (Lu et al., 2021).

## 3.3 Potential Moderators

Moderators help explain the conditions under which a particular relationship is strengthened, weakened, or varies across different contexts. Among the 36 articles we reviewed, only 7 included moderators that are relevant to at least one of cost of equity, cost of debt, and the weighted average cost of capital. This section addresses our third research question (RQ3), and all the moderators are summarized in Table 4.

**Heterogeneous investor beliefs**

Research indicates that when investors exhibit lower levels of heterogeneous beliefs, they are more likely to recognize and benefit from ESG performance, which garners greater interest from potential investors. Conversely, as heterogeneous investor beliefs increase, banks and other financial institutions usually adopt stringent credit policies to reduce risks, which further weakens the external financing ability of enterprises (Tang, 2022).

**Narrative risk disclosure quality**

Narrative risk disclosure quality is analyzed as a moderating variable in the study by Ismail and Obiedallah (2022). The author examined the moderating effect in their second hypothesis (H2): "Narrative risk disclosure quality has a significant impact on the association between firm performance and cost of equity capital in the Egyptian listed firms." Their finding supports the idea that, in the Egyptian capital market, the relationship between firm performance and the cost of equity is influenced by the quality of narrative risk disclosures in annual reports The results align with the assumptions of signaling theory, suggesting that providing high-quality risk disclosures, which enhance financial transparency, leads to a lower cost of equity.

**ESG rating**

Zhang et al (2022) investigated the ESG rating (Grade) by exploring the moderating effect of Grade between the data distribution of the monetary policy uncertainty (ShiborStd) and spread. The coefficient of the interaction is negative and statistically significant. This shows that companies with better ESG ratings may experience a reduced sensitivity of their spreads to fluctuations in monetary policy uncertainty.

**Financial distress**

Barrak et al (2023) investigated the moderating effect of financial distress between cost of debt and ESG. Their analysis across all models revealed that financial distress has a significant positive impact on borrowing costs. This finding indicates that companies experiencing greater financial distress tend to face higher expenses associated with debt financing.

**Firms' riskiness**

Srivastava et al (2022) proposed the moderating role of firms' riskiness on the CSR engagement and debt financing relationship during the pandemic. The authors use Default probability and Altman Z score as proxies for financial riskiness. They achieved a significant and positive coefficient of the interaction when the moderator is default probability. This implies CSR engagement may play a more pronounced role in influencing debt financing conditions for riskier firms.

**Sustainable development**

The moderator, sustainable development, has been examined from Sun et al (2023). Sustainable development is negatively related to financing costs and has weakened the inhibition of corporate environmental performance on financing costs (cost of debt). This implies that strong corporate environmental performance typically leads to lower financing costs, this effect is less pronounced in firms with a strong focus on sustainable development.

## 3.4 Endogeneity and Instrumental Variables

Endogeneity is a common concern in the field of corporate finance research (Roberts and Whited, 2013). Commonly adopted methods to cope with endogeneity include three main categories: generalized method of moment (GMM), instrument variables, and employment of fixed effect models. We summarize 7 instrumental variables utilized with respect to dealing with endogeneity concerns. These instruments are categorized by themes, with the first 3 incorporated in the cost of equity studies while the remaining 4 examined in the cost of debt papers. This section addresses our fourth research question (RQ4), and a detailed summary of these moderators is provided in Table 5.

**Corporate social responsibility and firm characteristic**

Governance performance disclosure and the cost of equity capital are interdependent, making it difficult to establish causality between the two. Furthermore, Other factors, such as firm size, industry characteristics, or economic conditions, influence both governance disclosure and the cost of equity, introducing bias if not properly controlled. Due to the endogenous relationship between governance performance disclosure and cost of equity capital, Mulchandani et al (2022) selected corporate social responsibility, firm size, leverage and ROA as instrument variables for ESG. Specifically, Ismail and Obiedallah (2022) utilized lagged firm size at time periods (t-1) and (t-2) as instrumental variables in their models to mitigate endogeneity issues.

**Key city of digital transformation**

The low cost of capital may drive enterprises to make digital transformation. In order to deal with possible endogenous problems, Hong et al (2023) used a dummy variable based on whether the enterprise is located in the key city of digital transformation. Enterprises situated in these key cities are expected to exhibit a higher degree of digital transformation. It is believed that there is a significant correlation between the city variable and the digital variable but there is no direct correlation with the equity capital cost. This ensures that the instrumental variable satisfies both the exogeneity and relevance conditions required for valid analysis.

**Country average corporate social performance**

To enhance the robustness of the study and address potential endogeneity arising from reverse causality or unobservable firm-specific variables (Garcia-Castro et al., 2010), the country average corporate social performance (CSP) score has also been used as an instrumental variable for firm-level CSP variable (Lynch & O'Hagan-Luff, 2023). The country-average CSP score is suitable as an instrumental variable as it is highly correlated with firm-level CSP, due to the fact that it is exposed to the same country-level factors that affect firm-level CSP and could only be associated with firm cost of capital through its impact on firm-level CSP.

**<u>Annual report comment letters (ARCL)</u>**

Previous studies have shown that annual report comment letters (ARCL) significantly increase a firm's cost of debt. However, firms with a high cost of debt are often associated with higher operational risks and lower profitability, which, in turn, increases the likelihood of receiving annual report comment letters (Cassell et al., 2013). To address the potential endogeneity between annual report comment letters and the cost of debt, Zhu et al. (2023) employed the natural logarithm of the distance between the company's location and the stock exchange (lndis) as an instrumental variable in a two-stage least squares (2SLS) regression.

**<u>Industry-level ESG measurement</u>**

Referring to Benlemlih and Bitar (2018), Kong (2023) employed the industry average ESG score and the enterprise initial ESG score as the instrument variables of ESG in his study. Similarly, the industry level measurement is used to estimate firm-specific sustainable index (Gao and Wan, 2023; Zhang et al., 2023). The average value of the ESG score of the industry, ruling out the score for the focal firm, has been considered as the instrument variable. A firm's scores are likely to have a correlation with scores of other firms within the same industry. Thus, the ESG rating of each enterprise will be affected by the ESG rating of other enterprises in the same industry, and the ESG rating of other enterprises is not directly related to the cost of equity capital of an enterprise (Quan et al., 2015; Dahiya and Singh, 2021; Tang, 2022).

**<u>Sustainable developed goals (SDG)</u>**

In the study by Ali et al. (2023), the observed results may be influenced by a common factor that simultaneously affects both the cost of debt and the measure of climate change action.

To cope with the situation that their observed negative relationship could merely be reflecting this endogeneity, they used the sustainable developed goals (SDG) dummy as the instrument variable for the climate action variable. This dummy variable is assigned a value of 1 if the Sustainable Development Goals have been established.

**Firm-level energy efficiency**

A fundamental econometric challenge when investigating the influence of a firm's energy efficiency on its cost of debt is possible endogeneity (Reddy and Sasidharan 2023). This arises from the possibility of reverse causation, where lower debt costs may incentivize firms to improve their energy efficiency rather than energy efficiency leading to reduced debt costs. Roy (2023) employed the "net financial expenses" as an instrument variable for energy efficiency. Additionally, energy expenses were used as an IV to tackle endogeneity concerns, as firms with high energy expenses may face challenges related to the scale of the economy.

## 3.5 Suggestions for Future Research

Table 6 lists the 15 articles that have suggested future research avenues. To answer our fifth research question (RQ5), we distill these reviewed articles and summarize their suggestions under five broader thematic categories.

**Methodological Advancements in ESG-CoC Research**

Tang (2022) suggested alternative models can be employed in future research, such as the implied cost of capital model (Gebhardt and Swaminathan, 2001). Additionally, two potential areas of investigation include: How other factors influence the relationship between ESG and the cost of equity capital, and whether analyst tracking can reveal this relationship and if risk-related factors can be used to explain it.

Zandi et al (2022) recommended future scholars to incorporate control variables, such as risk proxies, to better establish the relationship between the cost of equity (COE) capital and corporate governance. They also suggested using the weighted average cost of capital (WACC) as a more comprehensive measure of COE or exploring different ex-ante models to evaluate COE. Additionally, the interrelationship among the cost of debt, corporate governance, and the cost of equity capital, as highlighted by Alali et al. (2012) could be investigated in future research.

At very high levels, the incremental benefits of additional Corporate Social Performance efforts may diminish or even lead to overinvestment concerns, creating a non-linear effect. Lynch and O' Hagan-Luff (2023) suggested future research could investigate whether a complex non-linear relationship exists, influenced by heterogeneous investor preferences and varying incentives at different levels of Corporate Social Performance.

**Expanding Data and Measurement Approaches**

Rojo-Suarez and Alonso-Conde (2023) suggested future research should explore various aspects to better understand the effects of ESG performance on corporate value. This includes expanding the sample period under study and identifying alternative proxies for ESG variables that allow for the use of longer time series data. Future studies should also investigate the extent to which other value creation metrics, such as Economic Value Added (EVA), might reveal effects not considered in their research. It is also essential to seek exploration of the influence of ESG performance on borrowing capacity.

Mulchandani et al (2022) suggested researchers may build future work by incorporating other sustainability parameters. Further, exploring the impact of ESG disclosure on market capitalization and shareholder returns is recommended. The direct and indirect impact of ESG disclosure on specific financial metrics like market capitalization remains underexplored. Addressing these gaps can enhance the depth and scope of ESG research.

Ismail and Obiedallah (2022) highlighted the need for further research utilizing alternative sources of risk disclosure, such as board of directors' reports, financial releases, social media disclosures, and online corporate governance reports. Their study did not explore sub-classifications related to the tone and time orientation of the reported risk information. Consequently, investigating whether the tone of risk disclosures influences investor reactions in the capital market is a promising area for future research.

**Cross-Regional and Cross-Sectoral Considerations**

Mio et al (2023) suggested future research could expand by using a larger sample from multiple countries, alternative ESG databases and using other cost of equity measures. Further work may also explore whether the ESG rating disagreement affects the relationship between ESG performance and the cost of equity in the utility sector. Alternative business cases for ESG performance within the utility sector could also be explored to provide deeper insights.

Feng and Wu (2021) suggest that future research on the impacts of ESG disclosure on REIT capital structure, operations and performance should be fruitful and hold significant potential. Investors increasingly prioritize ESG factors when making investment decisions. Understanding how ESG disclosure affects REITs' capital structure and performance could help attract socially conscious investors and improve access to capital.

ESG disclosures may be evaluated by investors differently depending on the sector. Robust environmental disclosures might be particularly critical in sectors with high carbon footprints, affecting capital costs more significantly than in low-impact sectors. When using the panel-corrected standard error approach to explore whether ESG disclosure is relevant to cost of capital,

Kumawat and Patel (2022) did not consider sector-wise analysis, thus future research could address different sectors.

**Corporate Governance, Financial Structure, and ESG-CoC Dynamics**

Future research could explore how the different compositions of a company's owners and board members affect its debt costs (Thanatawee, 2023). The different compositions of a company's owners and board members may have an influence on governance and risk management, decision-making and strategic priorities, peputation and market perception. These can be further looked into.

Gigante and Manglaviti (2022) suggest future research may further investigate the potential difference in marginal effects of improved ESG performance on the cost of debt between the two subgroups. The firms can be divided into above-average group and below-average group companies from an ESG score point of view.

Teti et al (2022) suggest that future research should deepen the implications of non-market factors. Policy intervention and fiscal incentives may weight on the ability of issuer to achieve convenient cost of funding. Policy interventions, such as interest rate subsidies, green bonds, or tax credits, can directly impact the cost of funding for issuers. Understanding these effects can help identify how governments shape financial market support specific initiatives.

**ESG in Emerging Markets and Green Financing**

Lavin and Montecinos-Pearce (2022) suggest future researcher to investigate the aggregate effect on economic growth that ESGD produces in emerging economies that have less access to credit than firms in developed economies. Emerging economies often face underdeveloped financial markets. Investigating how ESGD influences economic growth could provide insights into whether ESG practices influence in less developed markets.

ESG practices are generally less established in emerging markets compared to developed economies. Agnese and Giacomini (2023) suggest future studies should extend the analysis to, particularly, emerging countries in order to ascertain whether bondholders price ESG practices in such contexts. Emerging countries often have distinct economic conditions, and regulatory frameworks, that may influence how ESG practices are perceived and valued by bondholders.

Greenium occurs when green bonds are priced higher than regular bonds, leading to lower returns for investment (Grishunin et al., 2023). Grishunin et al (2023) suggest that a higher greenium reduces long-term default risks for companies financed by banks. However, the resulting contradiction needs to be studied in detail in future research. Studying these contradictions in detail will help provide a more comprehensive understanding of the greenium-default risk relationship, ensuring that green financing achieves its intended benefits without introducing unintended risks or inefficiencies.

## 4. Final Thoughts

Synthesizing findings from multiple research studies can be an incredibly challenge task due to its lengthy duration and substantial financial resources required. Conducting a complex systematic literature review often cannot be completed within several months and may even take years. The most time-consuming steps are title screening and full-text screening. For instance, an article title may lack sufficient detail about the study, requiring researchers to having to review the abstract before making a decision during the title screening step.  In the full-text screening step, researchers must read at least a large portion of the article to evaluate its relevance and alignment with the pre-specified inclusion/exclusion criteria. To minimize biases, it typically requires at least two researchers working independently to determine whether a study meets the inclusion/exclusion criteria during both steps. Solving discrepancies between researchers can also be a time-intensive process. To enhance the efficiency of the title and full-text screening processes, AI-enabled assistant can be employed to learn the inclusion criteria by analyzing previous decisions.

In conclusion, our AI-assisted systematic literature review provides a comprehensive examination of the relationship between ESG and the cost of capital (CoC), demonstrating that strong ESG performance is associated with lower cost of capital with the incorporation of the pandemic and AI tools can significantly enhance the efficiency of title screening and aid statistical analysis. By systematically reviewing 36 published research articles, with the help of HubMeta platform, we answered our 5 research questions by synthesizing the theoretical frameworks employed, identifying key moderating factors, and highlighting the instrumental variables used to address endogeneity issues to fill the gap between ESG and cost of capital during the pandemic. Our study highlighted the stakeholder theory, agency theory, and risk-reward theory, which provides a useful framework for understanding the relationship between the cost of capital and firm ESG performance. The moderators that affect the relationship between cost of capital and ESG include heterogeneous investor beliefs, narrative risk disclosure quality, ESG rating, financial distress, firms' riskiness, and sustainable development. The instrument variables for solving endogeneity summarized in our study involve corporate social responsibility & firm characteristics, key city of digital transformation, country average corporate social performance, annual report comment letters (ARCL), industry-level ESG measurement, sustainable developed goals (SDG), and firm-level energy efficiency. Future research should focus on methodological advancements in ESG-CoC research, expanding data & measurement approaches, cross-regional & cross-sectoral considerations, corporate governance, financial structure, & ESG-CoC dynamics, and ESG in emerging markets & green financing. Our findings contribute to a deeper understanding of how

ESG practices influence the cost of capital and offer a foundation for further academic inquiry. These insights hold critical implications for investors and policymakers by showing the impact of an exogenous crisis and insurance-like mechanism of ESG for the cost of capital.

Determining a proper stopping point for machine learning is crucial when conducting an AI-assisted systematic literature review. Stopping too early may cause the AI assistant to reject papers that should not be excluded, lowering the accuracy and effectiveness of the entire process. Conversely, stopping too late can reduce the efficiency of title and full-text screenings, increasing the time and resources required to complete the systematic review. The AI-assistant errors can be categorized into two types: false positive and false negative. Minimizing false negative errors is particularly important. A higher false positive error rate means more irrelevant papers will progress to the full-text screening or final manual review stages. While this increases the workload, researchers still have the opportunity to identify and exclude irrelevant studies at later stages. However, a high false negative error rate results in relevant studies being excluded during the title or full-text screening steps. Once these studies are dropped, they cannot be re-included, leading to the loss of valuable articles.

When using AI-assisted tool to conduct systematic literature review, it is also important to note that outcomes of each machine learning attempt may vary. Even if the same prior decisions are used to train the AI in each attempt, the scores assigned to the same paper may differ slightly. In such cases, attention can be directed toward the ranking of each paper. If the ranking remains generally consistent, or if the positions of papers above and below the cutoff point do not change significantly, the results can still be considered valid. Furthermore, the size of the paper for AI training can vary based on the complexity of the inclusion/exclusion criteria, implying that a best/universal practice for training AI for the literature screening process is not feasible.

# Figures and Tables
## Figure 1
## Preferred Reporting Items for Systematic Reviews and Meta-Analyses (PRISMA)

This flowchart provides standardized guidelines to show our systematic reviews are reported completely and transparently. It indicates the number of literatures that remained in our project after the search stage, duplication, title screening, full-text screening, and meta-analysis. The numbers of the literature excluded due to inclusion/exclusion criteria are also reported in the flowchart.

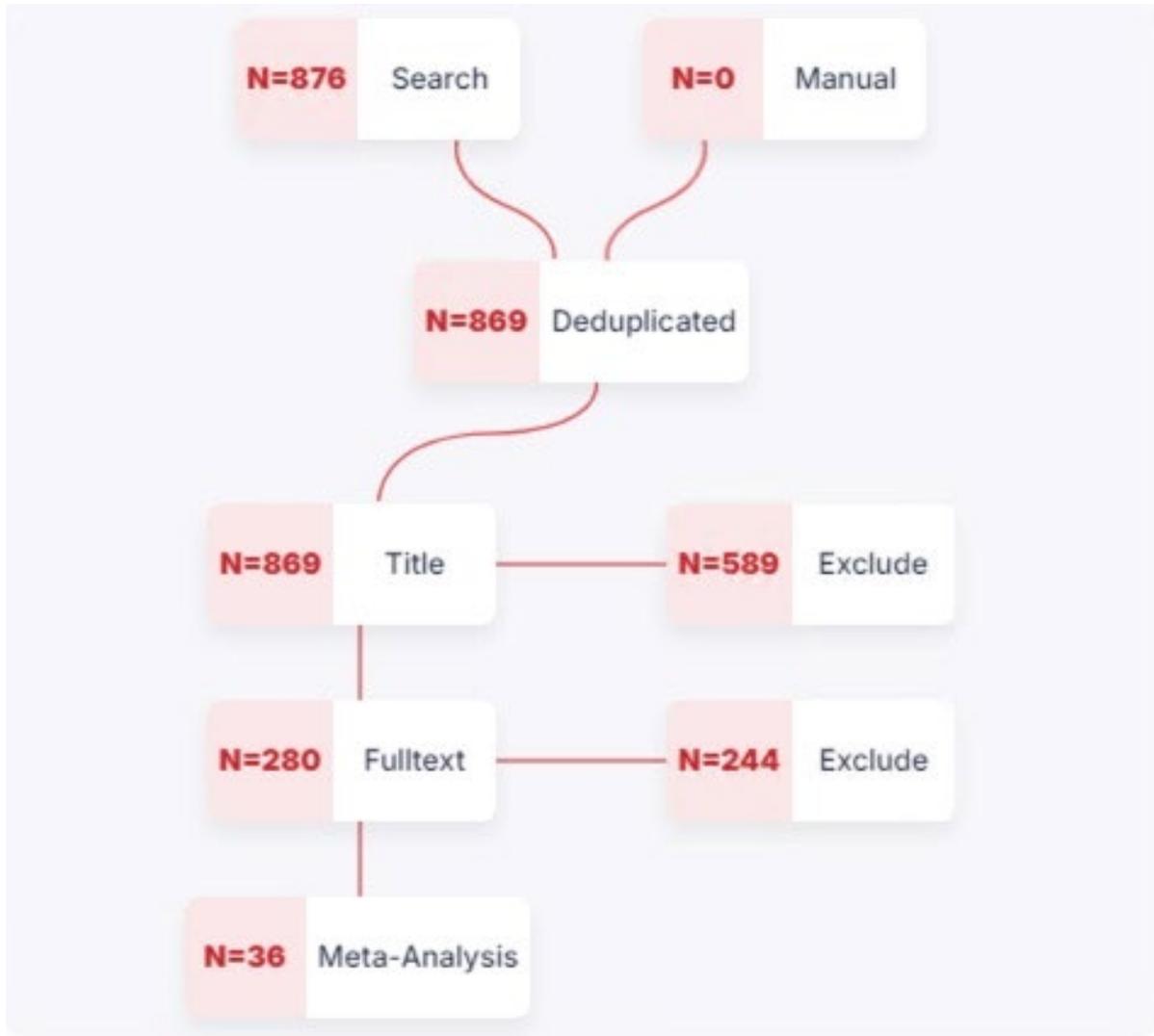

**Figure 2**
**Search keywords and Codes**

The syntax enabled Web of Science to locate studies in the field of ESG and the cost of capital published between 2020 and 2023. To identify appropriate keywords using ChatGPT, we prompted it with the following instruction: *I am conducting a systematic review on the influence of ESG on firms' cost of capital during the COVID-19 pandemic. Please review academic literature and suggest 20 relevant keywords that are most frequently mentioned in this context.* If the initial results included many irrelevant keywords, we refined the search by instructing ChatGPT: *Out of these keywords, only a few are relevant. Please suggest additional relevant keywords, focusing specifically on ESG and its influence on firms' cost of capital during the COVID-19 pandemic. Organize the results accordingly.* This iterative process ensured a comprehensive and relevant set of keywords for our research.

```
(TS=("capital structure" AND "ESG") OR TS=("cost of capital" AND "ESG")
OR TS=("cost of equity" AND "ESG") OR TS=("cost of debt" AND "ESG")
OR TS=("cost of capital" AND "environment") OR TS=("cost of equity" AND
"environment") OR TS=("cost of debt" AND "environment") OR TS=("cost
of capital" AND "social") OR TS=("cost of equity" AND "social") OR TS=
("cost of debt" AND "social") OR TS=("cost of capital" AND "governance")
OR TS=("cost of equity" AND "governance") OR TS=("cost of debt" AND
"governance") OR TS=("equity financing" AND "ESG") OR TS=("debt
financing" AND "ESG") OR TS=("financing" AND "ESG") OR TS=("equity
finance" AND "ESG") OR TS=("debt finance" AND "ESG") OR TS=("finance"
AND "ESG")) AND PY=2020-2023
```

**Figure 3**
**MSCI ESG Key Components Hierarchy**

The MSCI ESG Key Components Hierarchy provides a structured framework to understand and organize ESG-related topics. There are 10 themes under the 3 pillars and the 33 ESG Key issues help further narrow down specific topics. These create the checklist for us to determine whether to include or exclude the paper in the title screening or full-text screening stages.

| 3 Pillars | 10 Themes | 33 ESG Key Issues |
|---|---|---|
| Environmental | Climate Change | Carbon Emissions |
| | | Climate Change Vulnerability |
| | | Financing Environmental Impact |
| | | Product Carbon Footprint |
| | Natural Capital | Biodiversity & Land Use |
| | | Raw Material Sourcing |
| | | Water Stress |
| | Pollution & Waste | Electronic Waste |
| | | Packaging Material & Waste |
| | | Toxic Emissions & Waste |
| | Environmental Opportunities | Opportunities in Clean Tech |
| | | Opportunities in Green Building |
| | | Opportunities in Renewable Energy |
| Social | Human Capital | Health & Safety |
| | | Human Capital Development |
| | | Labor Management |
| | | Supply Chain Labor Standards |
| | Product Liability | Chemical Safety |
| | | Consumer Financial Protection |
| | | Privacy & Data Security |
| | | Product Safety & Quality |
| | | Responsible Investment |
| | Stakeholder Opposition | Community Relations |
| | | Controversial Sourcing |
| | Social Opportunities | Access to Finance |
| | | Access to Health Care |
| | | Opportunities in Nutrition & Health |

# Figure 4
## Refinitiv ESG Key Components Hierarchy

The MSCI ESG Key Components Hierarchy provides a structured framework to understand and organize ESG-related topics. There are 10 categories under the 3 pillars and the 25 ESG Key themes help further narrow down specific topics. These create the checklist for us to determine whether to include or exclude the paper in the title screening or full-text screening stages.

| Pillars | Categories | Themes | Data points | Weight method |
|---|---|---|---|---|
| Environmental | Emmission | Emissions | TR.AnalyticCO2 | Quant industry median |
| | | Waste | TR.AnalyticTotalWaste | Quant industry median |
| | | Biodiversity* | | |
| | | Environmental management systems* | | |
| | Innovation | Product innovation | TR.EnvProducts | Transparency weights |
| | | Green revenues, research and development (R&D) and capital expenditures (CapEx) | TR.AnalyticEnvRD | Quant industry median |
| | Resource use | Water | TR.AnalyticWaterUse | Quant industry median |
| | | Energy | TR.AnalyticEnergyUse | Quant industry median |
| | | Sustainable packaging* | | |
| | | Environmental supply chain* | | |
| Social | Community | Equally important to all industry groups, hence a median weight of five is assigned to all | | Equally important to all industry groups |
| | Human rights | Human rights | TR.PolicyHumanRights | Transparency weights |
| | Product responsibility | Responsible marketing | TR.PolicyResponsibleMarketing | Transparency weights |
| | | Product quality | TR.ProductQualityMonitoring | Transparency weights |
| | | Data privacy | TR.PolicyDataPrivacy | Transparency weights |
| | Workforce | Diversity and inclusion | TR.WomenEmployees | Quant industry median |
| | | Career development and training | TR.AvgTrainingHours | Transparency weights |
| | | Working conditions | TR.TradeUnionRep | Quant industry median |
| | | Health and safety | TR.AnalyticLostDays | Transparency weights |
| Governance | CSR strategy | CSR strategy | Data points in governance category and governance pillar | Count of data points in each governance category/all data points in governance pillar |
| | | ESG reporting and transparency | | |
| | Management | Structure (independence, diversity, committees) | Data points in governance category and governance pillar | Count of data points in each governance category/all data points in governance pillar |
| | | Compensation | | |
| | Shareholders | Shareholder rights | Data points in governance category and governance pillar | Count of data points in each governance category/all data points in governance pillar |
| | | Takeover defenses | | |

*No data points available that may be used as a proxy for ESG magnitude/materiality

**Figure 5**
**Group, Construct, and Measurements Hierarchical Breakdown**

Panel A and Panel B are visualized structures which shows the hierarchical breakdown of how cost of capital and ESG are categorized into three levels: Group Construct, Construct, and Measurements. The Group construct level shows the overarching concept. The Construct level shows the sub-concepts based on the group construct. Each construct is associated with specific research studies or methodologies that measure or analyze it.

**Panel A: Group, Construct, and Measurements regarding Cost of Capital**

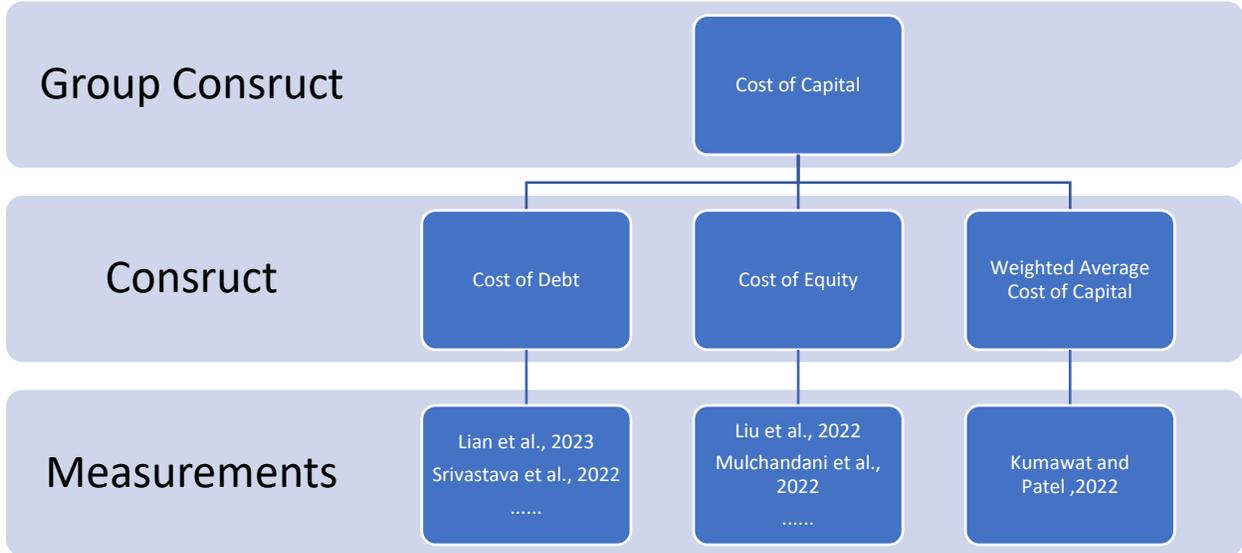

**Panel B: Group, Construct, and Measurements regarding ESG**

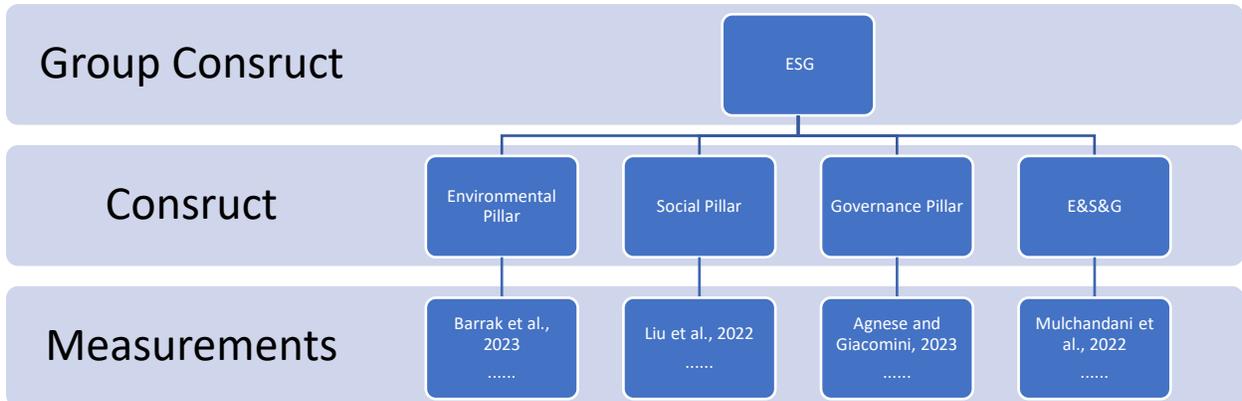

**Table 1: Summary of Papers Based on Relevance of COVID**

This table provides the list of the studies organized by cost of capital themes and the relevance of COVID/pandemic. Our systematic review examines 36 research papers. The papers are categorized into high relevance-group and low relevance-group. The cost of capital themes includes the cost of equity, the cost of debt, and the weighted average cost of capital.

| References | Relevance of COVID | | Themes | | |
|---|---|---|---|---|---|
| | High Relevance | Medium Relevance | COE | COD | WACC |
| Agnese and Giacomini, 2023 | ☑ | | | ○ | |
| Al Barrak et al., 2023 | | ☑ | | ○ | |
| Alekneviciene and Stralkute, 2023 | | ☑ | | ○ | |
| Ali et al., 2022 | ☑ | | | ○ | |
| Andries and Sprincean, 2023 | | ☑ | | ○ | |
| Arora and Sharma, 2022 | | ☑ | | ○ | |
| Chouaibi et al., 2022 | ☑ | | ○ | | |
| Chouaibi et al., 2021 | ☑ | | ○ | | |
| Feng and Wu, 2021 | | ☑ | | ○ | |
| Ferriani, 2023 | ☑ | | | ○ | |
| Gao and Wan, 2023 | | ☑ | | ○ | |
| Gigante and Manglaviti, 2022 | | ☑ | | ○ | |
| Grishunin et al., 2023 | | ☑ | | ○ | |
| Hong et al., 2023 | | ☑ | ○ | | |
| Ismail and Obiedallah, 2022 | | ☑ | ○ | | |
| Kong et al., 2023 | | ☑ | | ○ | |
| Kumawat and Patel, 2022 | | ☑ | | | ○ |
| Lavin and Montecinos-Pearce, 2022 | | ☑ | | ○ | |
| Lian et al., 2023 | ☑ | | | ○ | |
| Liu et al., 2022 | | ☑ | ○ | | |
| Lynch and O'Hagan-Luff, 2023 | | ☑ | ○ | | |
| Martellini and Vallee, 2021 | | ☑ | | ○ | |
| Mio et al., 2023 | | ☑ | ○ | | |
| Mulchandani et al., 2022 | | ☑ | ○ | | |
| Rojo-Suarez and Alonso-Conde, 2023 | ☑ | | ○ | | |
| Roy, 2023 | | ☑ | | ○ | |
| Srivastava et al., 2022 | ☑ | | | ○ | |
| Sun et al., 2023 | | ☑ | | ○ | |
| Tang, 2022 | | ☑ | ○ | | |
| Teti et al., 2022 | | ☑ | | ○ | |
| Thanatawee, 2023 | | ☑ | | ○ | |
| Wang et al., 2022 | | ☑ | | ○ | |

| Reference | Col1 | Col2 | Col3 |
|---|---|---|---|
| Zandi et al., 2022 | | ☑ | ○ |
| Zhang et al., 2023 | ☑ | | ○ |
| Zhang et al., 2022 | | ☑ | ○ |
| Zhu et al., 2023 | | ☑ | ○ |

**Table 2**
**Summary of Paper Based on Market**

This table demonstrates the studies organized by cost of capital themes and their corresponding market region. Our systematic review examines 36 research papers. The cost of capital themes includes the cost of equity, the cost of debt, and the weighted average cost of capital. The studies collect samples from African, American, Asian, European, Middle East, and multiple markets.

| References | Markets | | | | | | Themes | | |
|---|---|---|---|---|---|---|---|---|---|
| | African | American | Asian | European | Middle East | Multiple | COE | COD | WACC |
| Agnese and Giacomini, 2023 | | | | ☑ | | | | ○ | |
| Al Barrak et al., 2023 | | | | | ☑ | | | ○ | |
| Alekneviciene and Stralkute, 2023 | | | | ☑ | | | | ○ | |
| Ali et al., 2022 | | ☑ | | | | | | ○ | |
| Andries and Sprincean, 2023 | | | | | | ☑ | | ○ | |
| Arora and Sharma, 2022 | | | ☑ | | | | | ○ | |
| Chouaibi et al., 2022 | | | | | | ☑ | ○ | | |
| Chouaibi et al., 2021 | | | | ☑ | | | ○ | | |
| Feng and Wu, 2021 | | | | | | ☑ | | ○ | |
| Ferriani, 2023 | | | | | | ☑ | | ○ | |
| Gao and Wan, 2023 | | | ☑ | | | | | ○ | |
| Gigante and Manglaviti, 2022 | | | | ☑ | | | | ○ | |
| Grishunin et al., 2023 | | | | ☑ | | | | ○ | |
| Hong et al., 2023 | | | ☑ | | | | ○ | | |
| Ismail and Obiedallah, 2022 | ☑ | | | | | | ○ | | |
| Kong et al., 2023 | | | ☑ | | | | ○ | | |
| Kumawat and Patel, 2022 | | | ☑ | | | | | | ○ |
| Lavin and Montecinos-Pearce, 2022 | | ☑ | | | | | ○ | | |
| Lian et al., 2023 | | | ☑ | | | | | ○ | |
| Liu et al., 2022 | | | ☑ | | | | ○ | | |
| Lynch and O'Hagan-Luff, 2023 | | | | | | ☑ | ○ | | |

| Study | Col1 | Col2 | Col3 | Col4 |
|---|---|---|---|---|
| Martellini and Vallee, 2021 | | | ☑ | ○ |
| Mio et al., 2023 | | | ☑ | ○ |
| Mulchandani et al., 2022 | ☑ | | | ○ |
| Rojo-Suarez and Alonso-Conde, 2023 | | ☑ | | ○ |
| Roy, 2023 | ☑ | | | ○ |
| Srivastava et al., 2022 | | | ☑ | ○ |
| Sun et al., 2023 | ☑ | | | ○ |
| Tang, 2022 | ☑ | | ○ | |
| Teti et al., 2022 | | ☑ | | ○ |
| Thanatawee, 2023 | ☑ | | | ○ |
| Wang et al., 2022 | ☑ | | | ○ |
| Zandi et al., 2022 | ☑ | | ○ | |
| Zhang et al., 2023 | ☑ | | | ○ |
| Zhang et al., 2022 | ☑ | | | ○ |
| Zhu et al., 2023 | ☑ | | | ○ |

**Table 3**
**Summary of Paper Based on Theories**

This table provide the list of the studies organized by cost of capital themes and the most popular theories utilized in the studies. Our systematic review examines 36 research papers. The cost of capital themes includes the cost of equity, the cost of debt, and the weighted average cost of capital.

| References | Theories | | | | | | | | | | Themes | | |
|---|---|---|---|---|---|---|---|---|---|---|---|---|---|
| | Risk-Reward | Agency | Stakeholder | Signaling | Trade-off | Decision-usefulness | Legitimacy | Volunteer Disclosure | Resource Dependent | Peer Effect | COE | COD | WACC |
| Agnese and Giacomini, 2023 | ☑ | | | | | | | | | | | ○ | |
| Al Barrak et al., 2023 | | ☑ | ☑ | | | | | | | | | ○ | |
| Alekneviciene and Stralkute, 2023 | | | | ☑ | ☑ | | | | | | | ○ | |
| Ali et al., 2022 | | ☑ | ☑ | | | ☑ | | | | | | ○ | |
| Andries and Sprincean, 2023 | | | ☑ | | | | | | | | | ○ | |
| Arora and Sharma, 2022 | ☑ | | ☑ | | | | | | | | | ○ | |
| Chouaibi et al., 2022 | | | | ☑ | | | ☑ | | | | ○ | | |
| Chouaibi et al., 2021 | | | ☑ | | | | | | | | ○ | | |
| Feng and Wu, 2021 | | ☑ | | | | | | | | | | ○ | |
| Ferriani, 2023 | ☑ | | | | | | | | | | | ○ | |
| Gao and Wan, 2023 | | | ☑ | | | | | | | | | ○ | |
| Gigante and Manglaviti, 2022 | | ☑ | | | | | | | | | | ○ | |
| Grishunin et al., 2023 | ☑ | | | | | | | | | | | ○ | |
| Hong et al., 2023 | ☑ | | | | | | | | | | ○ | | |
| Ismail and Obiedallah, 2022 | | ☑ | | ☑ | | | | | | | ○ | | |
| Kong et al., 2023 | | | ☑ | | | | | | | | | ○ | |
| Kumawat and Patel, 2022 | | | | | | | ☑ | ☑ | | | | | ○ |
| Lavin and Montecinos-Pearce, 2022 | | ☑ | ☑ | | | | | | | | ○ | | |
| Lian et al., 2023 | ☑ | ☑ | | ☑ | | | | | ☑ | | | ○ | |
| Liu et al., 2022 | ☑ | | | | | | | | | | ○ | | |
| Lynch and O'Hagan-Luff, 2023 | | | ☑ | | | | | | | | ○ | | |
| Martellini and Vallee, 2021 | ☑ | | | | | | | | | | | ○ | |
| Mio et al., 2023 | ☑ | | | | | | | | | | ○ | | |
| Mulchandani et al., 2022 | | | ☑ | | | | ☑ | | | | ○ | | |
| Rojo-Suarez and Alonso-Conde, 2023 | | | ☑ | | | | | | | | ○ | | |
| Roy, 2023 | | | ☑ | | | | | | | | ○ | | |
| Srivastava et al., 2022 | | | | ☑ | | | | | | | ○ | | |
| Sun et al., 2023 | | ☑ | | ☑ | | | | | | | ○ | | |
| Tang, 2022 | | | ☑ | ☑ | | | | | | | ○ | | |
| Teti et al., 2022 | ☑ | | | | | | | | | | ○ | | |
| Thanatawee, 2023 | | | | | | | | | | ☑ | | ○ | |
| Wang et al., 2022 | | | | | | | | | | | | ○ | |
| Zandi et al., 2022 | | ☑ | | | | | | | | | ○ | | |
| Zhang et al., 2023 | | ☑ | | | | | | | | | | ○ | |
| Zhang et al., 2022 | ☑ | | | | | | | | | | | ○ | |
| Zhu et al., 2023 | | ☑ | | | | | | | | | | ○ | |

**Table 4**
**Summary of Paper Based on Moderators**

This table demonstrates the studies organized by cost of capital themes and the moderators utilized in the studies. Our systematic review examines 36 research papers. The cost of capital themes includes the cost of equity, the cost of debt, and the weighted average cost of capital.

| References | Moderators | | | | | | Themes | | |
|---|---|---|---|---|---|---|---|---|---|
| | Heterogeneous Investor Beliefs | Narrative Risk Disclosure Quality | ESG Rating | Financial Distress | Firms' Riskiness | Sustainable Development | COE | COD | WACC |
| Al Barrak et al., 2023 | | | | ☑ | | | | ○ | |
| Ismail and Obiedallah, 2022 | | ☑ | | | | | ○ | | |
| Srivastava et al., 2022 | | | | | ☑ | | | ○ | |
| Sun et al., 2023 | | | | | | ☑ | | ○ | |
| Tang, 2022 | ☑ | | | | | | ○ | | |
| Zhang et al., 2022 | | | ☑ | | | | | ○ | |

**Table 5**
**Summary of paper based on Instrument Variables**

This table shows the studies organized by cost of capital themes and the instrument variables. Our systematic review examines 36 research papers. The cost of capital themes includes the cost of equity, the cost of debt, and the weighted average cost of capital.

| References | Instrument Variables | | | | | | | Themes | | |
|---|---|---|---|---|---|---|---|---|---|---|
| | Corporate social responsibility and firm characteristics | Key city of digital transformation | Country average corporate social performance | Annual report comment letters (ARCL) | Industry-level ESG measurement | Sustainable developed goals (SDG) | Firm-level energy efficiency | COE | COD | WACC |
| Ali et al., 2022 | | | | | | ☑ | | ○ | | |
| Gao and Wan, 2023 | | | | | ☑ | | | ○ | | |
| Hong et al., 2023 | | ☑ | | | | | | ○ | | |
| Ismail and Obiedallah, 2022 | ☑ | | | | | | | ○ | | |
| Kong et al., 2023 | | | | | ☑ | | | | ○ | |
| Lynch and O'Hagan-Luff, 2023 | | | ☑ | | | | | ○ | | |
| Mulchandani et al., 2022 | ☑ | | | | | | | ○ | | |
| Roy, 2023 | | | | | | | ☑ | ○ | | |
| Zhang et al., 2023 | | | | | ☑ | | | ○ | | |
| Zhu et al., 2023 | | | | ☑ | | | | ○ | | |

# Table 6
# Summary of paper based on Future Research Avenue

This table shows the studies organized by the cost of capital themes and future research avenues. Our systematic review examines 36 research papers. The cost of capital themes includes the cost of equity, the cost of debt, and the weighted average cost of capital.

| References | Future Research | | | | | Themes | | |
|---|---|---|---|---|---|---|---|---|
| | Methodological Advancements in ESG-CoC Research | Expanding Data and Measurement Approaches | Cross-Regional and Cross-Sectoral Considerations | Corporate Governance, Financial Structure, and ESG-CoC Dynamics | ESG in Emerging Markets and Green Financing | COE | COD | WACC |
| Agnese and Giacomini, 2023 | | | | | ☑ | | ○ | |
| Feng and Wu, 2021 | | | ☑ | | | | ○ | |
| Gigante and Manglaviti, 2022 | | | | ☑ | | | ○ | |
| Grishunin et al, 2023 | | | | | ☑ | | ○ | |
| Ismail and Obiedallah, 2022 | | ☑ | | | | ○ | | |
| Kumawat and Patel, 2022 | | | ☑ | | | | | ○ |
| Lavin and Montecinos-Pearce, 2022 | | | | | ☑ | | ○ | |
| Lynch and O' Hagan-Luff, 2023 | ☑ | | | | | ○ | | |
| Mio et al., 2023 | | | ☑ | | | ○ | | |
| Mulchandani et al., 2022 | | ☑ | | | | ○ | | |
| Rojo-Suarez and Alonso-Conde, 2023 | | ☑ | | | | ○ | | |
| Tang, 2022 | ☑ | | | | | ○ | | |
| Teti et al., 2022 | | | | ☑ | | | ○ | |
| Thanatawee, 2023 | | | | ☑ | | | ○ | |
| Zandi et al., 2022 | ☑ | | | | | ○ | | |